# Structure and Composition of the 200 K-Superconducting Phase of H₂S under Ultrahigh Pressure: The Perovskite (SH⁻)(H₃S⁺)

*Elijah E. Gordon$^§$, Ke Xu$^§$, Hongjun Xiang, Annette Bussmann-Holder, Reinhard K. Kremer, Arndt Simon\*, Jürgen Köhler\* and Myung-Hwan Whangbo\**

**Abstract:** $H_2S$ is converted under ultrahigh pressure (> 110 GPa) to a metallic phase that becomes superconducting with a record $T_c$ of ~200 K. It has been proposed that the superconducting phase is body-centered cubic $H_3S$ ($Im\bar{3}m$, $a = 3.089$ Å) resulting from a decomposition reaction $3H_2S \rightarrow 2H_3S + S$. The analogy of $H_2S$ and $H_2O$ leads us to a very different conclusion. The well-known dissociation of water into $H_3O^+$ and $OH^-$ increases by orders of magnitude under pressure. An equivalent behavior of $H_2S$ is anticipated under pressure with the dissociation, $2H_2S \rightarrow H_3S^+ + SH^-$ forming a perovskite structure $(SH^-)(H_3S^+)$, which consists of corner-sharing $SH_6$ octahedra with $SH^-$ at each A-site (i.e., the center of each $S_8$ cube). Our DFT calculations show that the perovskite $(SH^-)(H_3S^+)$ is thermodynamically more stable than the $Im\bar{3}m$ structure of $H_3S$, and suggest that the A-site H atoms are most likely fluxional even at $T_c$.

For the phase of $H_2S$ under a pressure larger than 110 GPa high temperature superconductivity has been reported[1] with $T_c$ of *~200 K*. The decomposition of $H_2S$ into $SH_3 + S$ under such conditions has been deduced from synchrotron x-ray diffraction (XRD) experiments.[2,3,4] The obtained XRD patterns can be indexed based on a mixture of two phases, namely, the bcc $Im\bar{3}m$ structure ascribed to $H_3S$ with cell parameter $a = 3.089$ Å and the β-Po structure of sulfur.[3,5] Another XRD study under high pressure reported a more complex decomposition mixture that includes $H_3S$ and $H_4S_3$ phases.[6] DFT calculations confirmed the proposed $Im\bar{3}m$ structure of $H_3S$ as the lowest energy one.[2,7,8,9,10,11]

The $Im\bar{3}m$ structure of $H_3S$ ($a = 3.089$ Å) consists of two interpenetrating $SH_3$ perovskite sublattices (Figure 1a)) with a H···H

[a] Elijah E. Gordon, Prof. Dr. Myung-Hwan Whangbo
Department of Chemistry, North Carolina State University, Raleigh, NC 27695-8204 (USA)
Email: mike_whangbo@ncsu.edu

[b] Ke Xu, Prof. Dr. Hongjun Xiang
Key Laboratory of Computational Physical Sciences (Ministry of Education), State Key Laboratory of Surface Physics, and Department of Physics, Fudan University, Shanghai 200433, P. R. China
Collaborative Innovation Center of Advanced Microstructures, Nanjing 210093, P. R. China

[c] Prof. Dr. Annette Bussmann-Holder, Dr. Reinhard K. Kremer, Prof. Dr. Jürgen Köhler, Prof. Dr. A. Simon
Max-Planck-Institut für Festkörperforschung
Heisenbergstrasse 1, D-70569 Stuttgart (Germany)
Email: j.koehler@fkf.mpg.de
        a.simon@fkf.mpg.de

§ These authors contributed equally.

Supporting information for this article can be found under http://dx.doi.org/10.1002/anie.201511347.

contact distance of ~1.5 Å, which is very short compared with the van der Waals radii sum of 2.4 Å but very long compared with the H–H single-bond distance 0.74 Å in $H_2$. The decomposition, $3 H_2S \rightarrow 2 H_3S + S$, implies that the XRD intensities of $H_3S$ and sulfur with β-Po structure should have a 2:1 ratio. However, the comparison of the simulated XRD patterns with the observed ones taken from the literature in Figure S1 of the supporting information (SI) shows that the amounts of S in the samples vary and are always substantially smaller or hardly detectable than the expected one. Except for some spurious reflections the XRD patterns belong to a bcc lattice of sulfur atoms, but are not adequate enough to refine and assign the hydrogen atom positions with the Rietveld method. The bcc pattern is plotted in red color in Figure S1. The diagram in green color corresponds to sulfur with β-Po structure that should form in the decomposition reaction $3 H_2S \rightarrow 2H_3S + S$, however, sulfur does not show up except for one weak reflection at $2\Theta = 12°$. Thus the formation of $H_3S$ itself is questionable. The above-mentioned inconsistencies together with the apparent analogies between $H_2S$ and $H_2O$ were our motivation to reinvestigate the structure of the superconducting phase obtained from $H_2S$ under ultrahigh pressure.

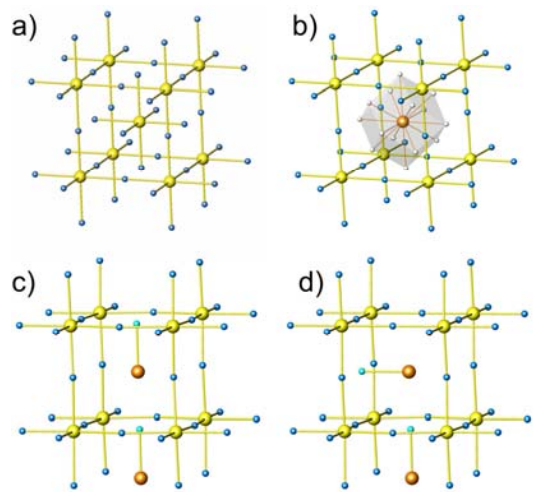

**Figure 1.** a) The proposed bcc structure of $H_3S$ ($Im\bar{3}m$, $a = 3.089$ Å) emphasizing the presence of two interpenetrating $SH_3$ perovskite sublattices. b) Structure of the perovskite $(SH^-)(H_3S^+)$ at ultrahigh pressure. The centering grey polyhedron represents the 14 orientations for the S–H bond (i.e., eight along the 3-fold rotational axes plus six along the 4-fold rotational axes. Only one of the 14 positions for the H atom of the $SH^-$ unit is occupied. c) The structure of $(SH^-)(H_3S^+)$ (*P4mm*, SG = 99) with A-site S–H bonds pointing to a face of the $S_8$ cube. d) The structure of $(SH^-)(H_3S^+)$ in which every two adjacent S–H bonds are perpendicular to each other (see text). Large yellow and orange circles correspond to the S atoms, and small blue circles to the H atoms.

At ambient conditions $H_2O$ dissociates into $H_3O^+$ and $OH^-$ to the extent given by $[H_3O^+][OH^-] = 10^{-14}$, which increases by orders of magnitude under high pressure. [2, 12] For $H_2S$ an analogous dissociation process for $H_2S$ into $H_3S^+$ and $SH^-$ under ultrahigh pressure can be expected, consistent with earlier ab initio calculations.[13] A perovskite structure $(SH^-)(H_3S^+)$ with nominal composition $H_2S$ and disordered $SH^-$ units at each A-site (i.e., the center of each $S_8$ cube) appears in reasonable agreement with the observed cubic lattice constant, see Figure 1 b). On the basis of DFT calculations, [14,15,16] we show that the perovskite $(SH^-)(H_3S^+)$ is thermodynamically more favorable than $H_3S$. In view of the expected dynamic motions of hydrogen, one can imagine that the functionalities of the S atoms on the A and B sites of the perovskite $ABO_3$ can easily be interchanged, according to $(SH^-)(H_3S^+) \Leftrightarrow (H_3S^+)(SH^-)$. In addition, our calculations reveal that at every A-site of a perovskite $(SH^-)(H_3S^+)$ the S–H bonds orient preferentially toward *any* one of the six faces of the $S_8$ cube, as shown in Figures 1c) and d). In case of all A-site S-H bonds pointing in one direction, the optimization of the crystal structure of $(SH^-)(H_3S^+)$ results in an arrangement of four H atoms of the $H_3S^+$ sublattice with short H···H contacts. These H atoms are displaced away from the H atom of the S–H bond (Figure 1c) (for the cif files of this *P4mm* structure, see SI). Alternatively, neighboring S–H units can be arranged orthogonal to each other (Fig 1d). There exists an infinite number of ways of arranging adjacent S–H bonds with 90° angles. Examples of some planar patterns of orthogonal S–H···S bonds are illustrated in Figure S2. They have small but significant energy differences, and the geometry optimization for $(SH^-)(H_3S^+)$ with a staggered pattern of zigzag chains (Figure 2) leads to an *Ima*2 (SG = 46) structure in which the H-S bonds of 1.492 and 1.597 Å alternate. This structure is the most stable one among the optimized structures with the planar patterns of Figure S2, see Table S1 (for the cif files of the optimized *Cmmm*, *Cmc*2₁ and *Ima*2 structures). It is important to emphasize that these structure optimizations rely on a static picture whereas in reality the hydrogen atoms are highly mobile and tunnel from side to side and carry out a hopping motion through the bcc sulfur lattice. Such a mobile configurational change may even lead to a plastic-like behavior, which would result in similar conclusions as derived here.[17]

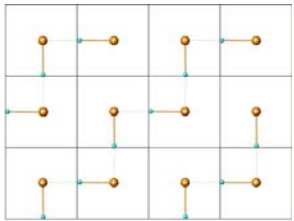

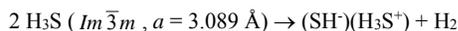

*Figure 2*. Schematic representation of a plane of orthogonally arranged S–H···S bridges made up of the A-site S–H bonds forming zigzag chains in $(SH^-)(H_3S^+)$. Blue circles represent the H atoms, orange circles S atoms.

We then estimate the relative stabilities of the $H_3S$ ($Im\overline{3}m$) and $(SH^-)(H_3S^+)$ structures. Since the compositions of the two phases are not identical, we first examine the enthalpy change ΔH of the hypothetical reaction,

2 $H_3S$ ($Im\overline{3}m$, a = 3.089 Å) → $(SH^-)(H_3S^+)$ + $H_2$

The calculations show that ΔH = -122.5, -170.9 and -194.3 meV, for the *Cmmm*, *Cmc*2₁ and *Ima*2 structures of $(SH^-)(H_3S^+)$, respectively, i.e., the $(SH^-)(H_3S^+)$ phase is energetically more stable than the $H_3S$ structure ($Im\overline{3}m$). In terms of the free energy ΔG = ΔH – TΔS, the $(SH^-)(H_3S^+)$ phase is even more favored. Because there are three possible orthogonal arrangements of S–H···S bridges at any given A-site, fluxional behavior of the A-site H atoms as mentioned above will lead to a configuration entropy gain ΔS = Rln3 = 2.18 cal/mol/deg. At 200 K this contributes -TΔS ≈ -18.9 meV to the free energy, so that the *Cmmm*, *Cmc*2₁ and *Ima*2 structures of $(SH^-)(H_3S^+)$ are more likely than the $H_3S$ structure ($Im\overline{3}m$) by ΔG = -140.4, -189.8 and -213.2 meV, respectively.

The calculated electronic structure of $(SH^-)(H_3S^+)$ with zigzag chains of S-H bonds (Figure 2) is presented in Figure 3. The occupied S 3p states of the B-site sulfur atoms (i.e., the atoms of the perovskite framework) lie lower in energy than those of the A-site sulfur atoms, which dominate the states at the Fermi level. The electronic structure description $(SH^-)(H_3S^+)$ is consistent with the formal charges. The flat band part along the Γ-X direction observed at the Fermi level resembles closely the corresponding band structure of the superconductor $MgB_2$ [18][19] for which a multigap superconductivity has been proposed. The Fermi surface exhibits cube-like features (see Figure S3 of the SI).

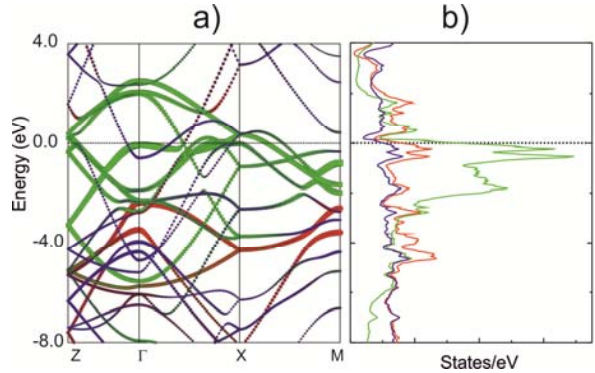

*Figure 3*. The electronic structure calculated for the ultrahigh pressure phase of $H_2S$ with the perovskite structure $(SH^-)(H_3S^+)$ containing zigzag chains of $SH^-$ units (SG = *Ima*2), s. Figure 2. a) The band dispersion relations with fat-band representations, where the 3p states of the A-site S atoms are represented in green, those of the B-site S atoms in red, and the s/p states of the H atoms in blue. b) The PDOS plot for the 3p states of the A-site S atoms (green), that of the B-site S atoms (red), and the s/p states of the H atoms (blue).

The remarkably high $T_c$'s in the 80–200 K range for the proposed $H_3S$ under high pressure have been explained on the basis of the strong-coupling theory of superconductivity, [7,8] with the electron-phonon coupling increasing strongly with increasing pressure where more than 90 % of the coupling arises from H vibrations. [9,10, 20] This is supported by the observation of a substantial isotope effect upon deuteration. However, the $T_c$ of the deuterated sample $D_2S$ is reduced to ~90 K at the highest pressure corresponding to an isotope exponent α ≈ 1, and this exponent varies with pressure.[21] Recently, Bianconi and Jahlborg, [22] showed that these observations are not explained by the single gap superconductivity theory of Eliashberg [23] and Allen-Dynes,[24] and pointed out the need to consider a multigap superconductivity theory.

In summary, concerning the high-$T_c$ superconducting phase of $H_2S$ under high pressure, our work presents a picture different from the currently adopted one. Namely, $H_2S$ does not decompose into $H_3S$ and S, but dissociates under the formation of a perovskite type structure $(SH^-)(H_3S^+)$. Furthermore, our work suggests that the A-site H atoms are most likely fluxional even at $T_c$.

## Experimental Section

We carried out non-spin-polarized DFT calculations employing the projector augmented wave method encoded in the Vienna ab initio simulation package[14-16] and the generalized gradient approximation of Perdew, Burke and Ernzerhof for the exchange-correlation functionals. [16] The plane wave cutoff energy of 1200 eV and the threshold of self-consistent-field energy convergence of $10^{-8}$ eV were used. The Brillouin zone was sampled by a set of 12×12×12 k-points.

## Acknowledgements

This research used resources of the National Energy Research Scientific Computing Center, a DOE Office of Science User Facility supported by the Office of Science of the U.S. Department of Energy under Contract No. DE-AC02-05CH11231. H.J.X. thanks NSFC and the Special Funds for Major State Basic Research.

**Keywords:** $H_2S$ superconductor · $H_2S$ perovskite · ionization under pressure · electronic structure

Supporting Information

for

**Structure and composition of the 200 K-superconducting phase of H$_2$S under ultrahigh pressure: the perovskite (SH$^-$)(H$_3$S$^+$)**

Elijah E. Gordon†[a], Ke Xu†[b], Hongjun Xiang[b], Annette Bussmann-Holder[c], Reinhard K. Kremer[c], Arndt Simon*[c], Jürgen Köhler*[c] and Myung-Hwan Whangbo*[a]

[1] Comparison of the experimental and simulated XRD patterns

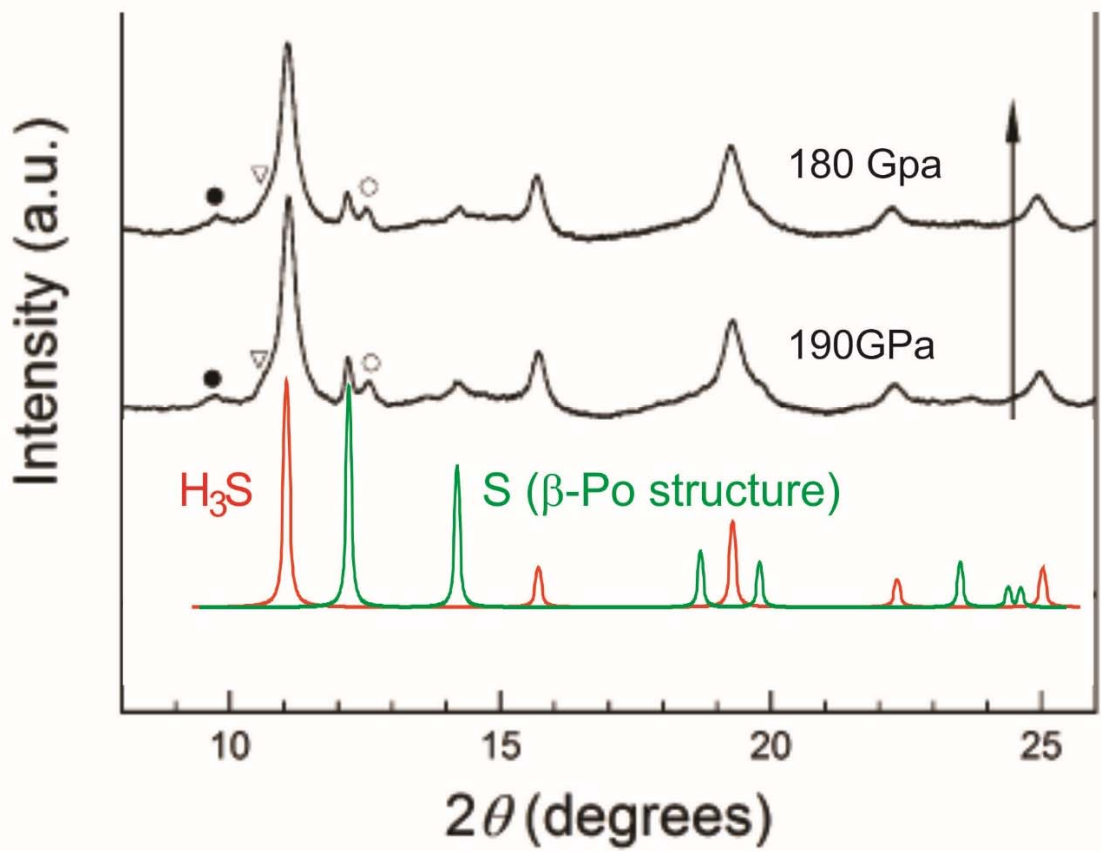

Figure 1. Experimental XRD patterns for D$_2$S at different pressures (taken from ref. S1) together with the calculated XRD patterns for the $Im\bar{3}m$ structure of D$_3$S (red curve) and β-Po elemental sulfur (green curve) at 180 GPa (taken from ref. S2, S3).

[2] **Fig. S2**

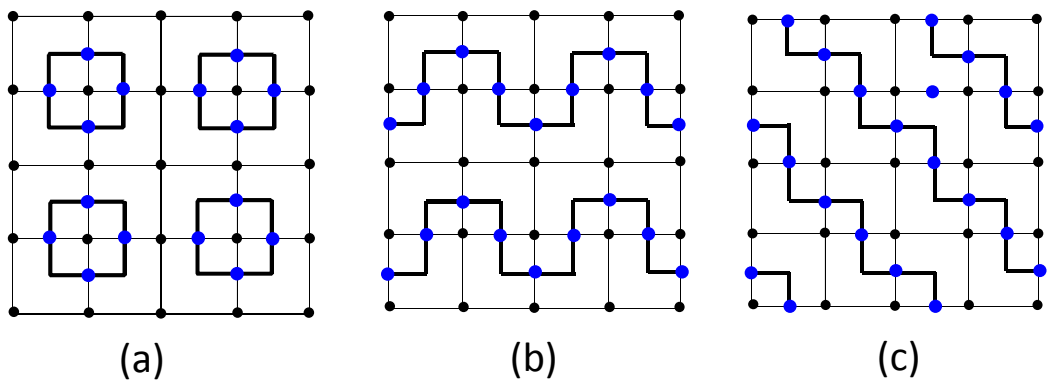

Fig. S2.    Three patterns of orthogonally-arranged S−H...S bridges made up of the A-site S−H bonds. (a) Isolated squares. (b) Meander chains. (c) Zigzag chains. The black circles represent the H atoms of the perovskite-framework, and the blue circles those of the A-site S−H bonds.

[3] **Table S1**

The relative energies per formula unit (FU), ΔE (meV/FU), of the $(SH^-)(H_3S^+)$ structures with the three different arrangements of the A-site S–H...S bridges depicted in **Fig. S2** as well as the P4mm structure of $(SH^-)(H_3S^+)$ shown in **Fig. 1c** are listed in **Table S1**. Also summarized in this table are the S–H bond lengths (Å) of the S–H...S bonds between A-site S atoms and the short H···H contacts (Å) between the A-site H and the B-site H atoms in the perovskite structures $(SH^-)(H_3S^+)$. The stabilities of the four structures increase in the order, P4mm < Cmmm < $Cmc2_1$ < Ima2. **Table S1** shows that the Ima2 structure with zigzag chain patterns of the S–H...S bridges is the most stable one, but the $Cmc2_1$ structure with meander chains is only slightly less stable than the Ima2 structure (by 23.4 meV/FU). Thermodynamically, these two structures are nearly equally probable around 200 – 300 K, given the associated thermal energies RT = 17.2 – 25.8 meV.

| Structure | P4mm SG = 99 | Cmmm SG = 65 | $Cmc2_1$ SG = 36 | Ima2 SG = 46 |
|---|---|---|---|---|
| Structural features | no orthogonal S–H...S | isolated squares | ribbon chains | zigzag chains |
| ΔE (meV/FU) | 383.0 | 71.8 | 23.4 | 0.0 |
| S–H–S | 1.338/1.751 | 1.545/1.544 | 1.498/1.591 1.541/1.548 | 1.492/1.597 |
| H···H contacts | 1.592 (×4) | 1.548 (×2) 1.547 (×2) 1.542 1.541 | 1.677 1.645 1.552 1.546 | 1.678 (×2) 1.645 (×2) |

[4] The cif files for the optimized structures of (SH$^-$)(SH$_3^+$).

(a) P4mm structure

#====================================================================
# CRYSTAL DATA
#--------------------------------------------------------------------

data_VESTA_phase_1

_pd_phase_name                  'Ž'
_cell_length_a                   6.17800
_cell_length_b                   6.17800
_cell_length_c                   6.17800
_cell_angle_alpha                90
_cell_angle_beta                 90
_cell_angle_gamma                90
_symmetry_space_group_name_H-M   'P 1'
_symmetry_Int_Tables_number      1

loop_
_symmetry_equiv_pos_as_xyz
 'x, y, z'

loop_
  _atom_site_label
  _atom_site_occupancy
  _atom_site_fract_x
  _atom_site_fract_y
  _atom_site_fract_z
  _atom_site_adp_type
  _atom_site_B_iso_or_equiv
  _atom_site_type_symbol
   S1    1.0   0.000000   0.000000   0.000000   Biso  1.000000 S
   S2    1.0   0.000000   0.000000   0.500000   Biso  1.000000 S
   S3    1.0   0.000000   0.500000   0.000000   Biso  1.000000 S
   S4    1.0   0.000000   0.500000   0.500000   Biso  1.000000 S
   S5    1.0   0.500000   0.000000   0.000000   Biso  1.000000 S
   S6    1.0   0.500000   0.000000   0.500000   Biso  1.000000 S
   S7    1.0   0.500000   0.500000   0.000000   Biso  1.000000 S
   S8    1.0   0.500000   0.500000   0.500000   Biso  1.000000 S

| | | | | | | |
|---|---|---|---|---|---|---|
| S9  | 1.0 | 0.250000 | 0.250000 | 0.250000 | Biso | 1.000000 S |
| S10 | 1.0 | 0.250000 | 0.250000 | 0.750000 | Biso | 1.000000 S |
| S11 | 1.0 | 0.250000 | 0.750000 | 0.250000 | Biso | 1.000000 S |
| S12 | 1.0 | 0.250000 | 0.750000 | 0.750000 | Biso | 1.000000 S |
| S13 | 1.0 | 0.750000 | 0.250000 | 0.250000 | Biso | 1.000000 S |
| S14 | 1.0 | 0.750000 | 0.250000 | 0.750000 | Biso | 1.000000 S |
| S15 | 1.0 | 0.750000 | 0.750000 | 0.250000 | Biso | 1.000000 S |
| S16 | 1.0 | 0.750000 | 0.750000 | 0.750000 | Biso | 1.000000 S |
| H1  | 1.0 | 0.230390 | 0.000000 | 0.000000 | Biso | 1.000000 H |
| H2  | 1.0 | 0.230390 | 0.000000 | 0.500000 | Biso | 1.000000 H |
| H3  | 1.0 | 0.230390 | 0.500000 | 0.000000 | Biso | 1.000000 H |
| H4  | 1.0 | 0.230390 | 0.500000 | 0.500000 | Biso | 1.000000 H |
| H5  | 1.0 | 0.730390 | 0.000000 | 0.000000 | Biso | 1.000000 H |
| H6  | 1.0 | 0.730390 | 0.000000 | 0.500000 | Biso | 1.000000 H |
| H7  | 1.0 | 0.730390 | 0.500000 | 0.000000 | Biso | 1.000000 H |
| H8  | 1.0 | 0.730390 | 0.500000 | 0.500000 | Biso | 1.000000 H |
| H9  | 1.0 | 0.029044 | 0.250000 | 0.000000 | Biso | 1.000000 H |
| H10 | 1.0 | 0.029044 | 0.250000 | 0.500000 | Biso | 1.000000 H |
| H11 | 1.0 | 0.029044 | 0.750000 | 0.000000 | Biso | 1.000000 H |
| H12 | 1.0 | 0.029044 | 0.750000 | 0.500000 | Biso | 1.000000 H |
| H13 | 1.0 | 0.529044 | 0.250000 | 0.000000 | Biso | 1.000000 H |
| H14 | 1.0 | 0.529044 | 0.250000 | 0.500000 | Biso | 1.000000 H |
| H15 | 1.0 | 0.529044 | 0.750000 | 0.000000 | Biso | 1.000000 H |
| H16 | 1.0 | 0.529044 | 0.750000 | 0.500000 | Biso | 1.000000 H |
| H17 | 1.0 | 0.029044 | 0.000000 | 0.250000 | Biso | 1.000000 H |
| H18 | 1.0 | 0.029044 | 0.000000 | 0.750000 | Biso | 1.000000 H |
| H19 | 1.0 | 0.029044 | 0.500000 | 0.250000 | Biso | 1.000000 H |
| H20 | 1.0 | 0.029044 | 0.500000 | 0.750000 | Biso | 1.000000 H |
| H21 | 1.0 | 0.529044 | 0.000000 | 0.250000 | Biso | 1.000000 H |
| H22 | 1.0 | 0.529044 | 0.000000 | 0.750000 | Biso | 1.000000 H |
| H23 | 1.0 | 0.529044 | 0.500000 | 0.250000 | Biso | 1.000000 H |
| H24 | 1.0 | 0.529044 | 0.500000 | 0.750000 | Biso | 1.000000 H |
| H25 | 1.0 | 0.466592 | 0.250000 | 0.250000 | Biso | 1.000000 H |
| H26 | 1.0 | 0.466592 | 0.250000 | 0.750000 | Biso | 1.000000 H |
| H27 | 1.0 | 0.466592 | 0.750000 | 0.250000 | Biso | 1.000000 H |
| H28 | 1.0 | 0.466592 | 0.750000 | 0.750000 | Biso | 1.000000 H |
| H29 | 1.0 | 0.966592 | 0.250000 | 0.250000 | Biso | 1.000000 H |
| H30 | 1.0 | 0.966592 | 0.250000 | 0.750000 | Biso | 1.000000 H |
| H31 | 1.0 | 0.966592 | 0.750000 | 0.250000 | Biso | 1.000000 H |
| H32 | 1.0 | 0.966592 | 0.750000 | 0.750000 | Biso | 1.000000 H |

(b) Cmmm structure

#======================================================================

# CRYSTAL DATA

#----------------------------------------------------------------------

data_VESTA_phase_1

_pd_phase_name                    ' S   H                '
_cell_length_a                     6.17800
_cell_length_b                     6.17800
_cell_length_c                     6.17800
_cell_angle_alpha                   90
_cell_angle_beta                    90
_cell_angle_gamma                   90
_symmetry_space_group_name_H-M     'P 1'
_symmetry_Int_Tables_number         1

loop_
_symmetry_equiv_pos_as_xyz
 'x, y, z'

loop_
  _atom_site_label
  _atom_site_occupancy
  _atom_site_fract_x
  _atom_site_fract_y
  _atom_site_fract_z
  _atom_site_adp_type
  _atom_site_B_iso_or_equiv
  _atom_site_type_symbol
   S1    1.0   0.000000   0.000000   0.000000   Biso  1.000000 S
   S2    1.0   0.000000   0.000000   0.500000   Biso  1.000000 S
   S3    1.0   0.000000   0.500000   0.000000   Biso  1.000000 S
   S4    1.0   0.000000   0.500000   0.500000   Biso  1.000000 S
   S5    1.0   0.500000   0.000000   0.000000   Biso  1.000000 S
   S6    1.0   0.500000   0.000000   0.500000   Biso  1.000000 S
   S7    1.0   0.500000   0.500000   0.000000   Biso  1.000000 S
   S8    1.0   0.500000   0.500000   0.500000   Biso  1.000000 S
   S9    1.0   0.250000   0.250000   0.250000   Biso  1.000000 S
   S10   1.0   0.250000   0.250000   0.750000   Biso  1.000000 S
   S11   1.0   0.250000   0.750000   0.250000   Biso  1.000000 S
   S12   1.0   0.250000   0.750000   0.750000   Biso  1.000000 S

| | | | | | | |
|---|---|---|---|---|---|---|
| S13 | 1.0 | 0.750000 | 0.250000 | 0.250000 | Biso | 1.000000 S |
| S14 | 1.0 | 0.750000 | 0.250000 | 0.750000 | Biso | 1.000000 S |
| S15 | 1.0 | 0.750000 | 0.750000 | 0.250000 | Biso | 1.000000 S |
| S16 | 1.0 | 0.750000 | 0.750000 | 0.750000 | Biso | 1.000000 S |
| H1 | 1.0 | 0.250000 | 0.000000 | 0.000000 | Biso | 1.000000 H |
| H2 | 1.0 | 0.250000 | 0.000000 | 0.500000 | Biso | 1.000000 H |
| H3 | 1.0 | 0.250000 | 0.500000 | 0.000000 | Biso | 1.000000 H |
| H4 | 1.0 | 0.250000 | 0.500000 | 0.500000 | Biso | 1.000000 H |
| H5 | 1.0 | 0.750000 | 0.000000 | 0.000000 | Biso | 1.000000 H |
| H6 | 1.0 | 0.750000 | 0.000000 | 0.500000 | Biso | 1.000000 H |
| H7 | 1.0 | 0.750000 | 0.500000 | 0.000000 | Biso | 1.000000 H |
| H8 | 1.0 | 0.750000 | 0.500000 | 0.500000 | Biso | 1.000000 H |
| H9 | 1.0 | 0.000000 | 0.250000 | 0.000000 | Biso | 1.000000 H |
| H10 | 1.0 | 0.000000 | 0.250000 | 0.500000 | Biso | 1.000000 H |
| H11 | 1.0 | 0.000000 | 0.750000 | 0.000000 | Biso | 1.000000 H |
| H12 | 1.0 | 0.000000 | 0.750000 | 0.500000 | Biso | 1.000000 H |
| H13 | 1.0 | 0.500000 | 0.250000 | 0.000000 | Biso | 1.000000 H |
| H14 | 1.0 | 0.500000 | 0.250000 | 0.500000 | Biso | 1.000000 H |
| H15 | 1.0 | 0.500000 | 0.750000 | 0.000000 | Biso | 1.000000 H |
| H16 | 1.0 | 0.500000 | 0.750000 | 0.500000 | Biso | 1.000000 H |
| H17 | 1.0 | 0.999907 | 0.020674 | 0.253947 | Biso | 1.000000 H |
| H18 | 1.0 | 0.000093 | 0.020674 | 0.746053 | Biso | 1.000000 H |
| H19 | 1.0 | 0.999907 | 0.479326 | 0.253947 | Biso | 1.000000 H |
| H20 | 1.0 | 0.000093 | 0.479326 | 0.746053 | Biso | 1.000000 H |
| H21 | 1.0 | 0.499907 | 0.979326 | 0.253947 | Biso | 1.000000 H |
| H22 | 1.0 | 0.500093 | 0.979326 | 0.746053 | Biso | 1.000000 H |
| H23 | 1.0 | 0.499907 | 0.520674 | 0.253947 | Biso | 1.000000 H |
| H24 | 1.0 | 0.500093 | 0.520674 | 0.746053 | Biso | 1.000000 H |
| H25 | 1.0 | 0.499953 | 0.250000 | 0.250586 | Biso | 1.000000 H |
| H26 | 1.0 | 0.249649 | 0.250000 | 0.999894 | Biso | 1.000000 H |
| H27 | 1.0 | 0.250351 | 0.750000 | 0.000106 | Biso | 1.000000 H |
| H28 | 1.0 | 0.000047 | 0.750000 | 0.749414 | Biso | 1.000000 H |
| H29 | 1.0 | 0.750351 | 0.250000 | 0.000106 | Biso | 1.000000 H |
| H30 | 1.0 | 0.500047 | 0.250000 | 0.749414 | Biso | 1.000000 H |
| H31 | 1.0 | 0.999953 | 0.750000 | 0.250586 | Biso | 1.000000 H |
| H32 | 1.0 | 0.749649 | 0.750000 | 0.999894 | Biso | 1.000000 H |

(c) Cmc2$_1$ structure

#======================================================================

# CRYSTAL DATA

#----------------------------------------------------------------------

data_VESTA_phase_1

_pd_phase_name                  '…S   H                '
_cell_length_a                  6.17800
_cell_length_b                  6.17800
_cell_length_c                  6.17800
_cell_angle_alpha                  90
_cell_angle_beta                   90
_cell_angle_gamma                  90
_symmetry_space_group_name_H-M     'P 1'
_symmetry_Int_Tables_number        1

loop_
_symmetry_equiv_pos_as_xyz
  'x, y, z'

loop_
  _atom_site_label
  _atom_site_occupancy
  _atom_site_fract_x
  _atom_site_fract_y
  _atom_site_fract_z
  _atom_site_adp_type
  _atom_site_B_iso_or_equiv
  _atom_site_type_symbol
   S1    1.0   0.000000   0.000000   0.000000   Biso  1.000000 S
   S2    1.0   0.000000   0.000000   0.500000   Biso  1.000000 S
   S3    1.0   0.000000   0.500000   0.000000   Biso  1.000000 S
   S4    1.0   0.000000   0.500000   0.500000   Biso  1.000000 S
   S5    1.0   0.500000   0.000000   0.000000   Biso  1.000000 S
   S6    1.0   0.500000   0.000000   0.500000   Biso  1.000000 S
   S7    1.0   0.500000   0.500000   0.000000   Biso  1.000000 S
   S8    1.0   0.500000   0.500000   0.500000   Biso  1.000000 S
   S9    1.0   0.250000   0.250000   0.250000   Biso  1.000000 S
   S10   1.0   0.250000   0.250000   0.750000   Biso  1.000000 S
   S11   1.0   0.250000   0.750000   0.250000   Biso  1.000000 S
   S12   1.0   0.250000   0.750000   0.750000   Biso  1.000000 S

| | | | | | | |
|---|---|---|---|---|---|---|
| S13 | 1.0 | 0.750000 | 0.250000 | 0.250000 | Biso | 1.000000 S |
| S14 | 1.0 | 0.750000 | 0.250000 | 0.750000 | Biso | 1.000000 S |
| S15 | 1.0 | 0.750000 | 0.750000 | 0.250000 | Biso | 1.000000 S |
| S16 | 1.0 | 0.750000 | 0.750000 | 0.750000 | Biso | 1.000000 S |
| H1 | 1.0 | 0.258747 | 0.020725 | 0.009549 | Biso | 1.000000 H |
| H2 | 1.0 | 0.241253 | 0.979275 | 0.509549 | Biso | 1.000000 H |
| H3 | 1.0 | 0.258747 | 0.479275 | 0.009549 | Biso | 1.000000 H |
| H4 | 1.0 | 0.241253 | 0.520725 | 0.509549 | Biso | 1.000000 H |
| H5 | 1.0 | 0.758747 | 0.979275 | 0.009549 | Biso | 1.000000 H |
| H6 | 1.0 | 0.741253 | 0.020725 | 0.509549 | Biso | 1.000000 H |
| H7 | 1.0 | 0.758747 | 0.520725 | 0.009549 | Biso | 1.000000 H |
| H8 | 1.0 | 0.741253 | 0.479275 | 0.509549 | Biso | 1.000000 H |
| H9 | 1.0 | 0.018889 | 0.250000 | 0.004028 | Biso | 1.000000 H |
| H10 | 1.0 | 0.981111 | 0.250000 | 0.504028 | Biso | 1.000000 H |
| H11 | 1.0 | 0.981358 | 0.750000 | 0.998506 | Biso | 1.000000 H |
| H12 | 1.0 | 0.018642 | 0.750000 | 0.498506 | Biso | 1.000000 H |
| H13 | 1.0 | 0.481358 | 0.250000 | 0.998506 | Biso | 1.000000 H |
| H14 | 1.0 | 0.518642 | 0.250000 | 0.498506 | Biso | 1.000000 H |
| H15 | 1.0 | 0.518889 | 0.750000 | 0.004028 | Biso | 1.000000 H |
| H16 | 1.0 | 0.481111 | 0.750000 | 0.504028 | Biso | 1.000000 H |
| H17 | 1.0 | 0.003818 | 0.978667 | 0.230822 | Biso | 1.000000 H |
| H18 | 1.0 | 0.996182 | 0.978667 | 0.730822 | Biso | 1.000000 H |
| H19 | 1.0 | 0.003818 | 0.521333 | 0.230822 | Biso | 1.000000 H |
| H20 | 1.0 | 0.996182 | 0.521333 | 0.730822 | Biso | 1.000000 H |
| H21 | 1.0 | 0.503818 | 0.021333 | 0.230822 | Biso | 1.000000 H |
| H22 | 1.0 | 0.496182 | 0.021333 | 0.730822 | Biso | 1.000000 H |
| H23 | 1.0 | 0.503818 | 0.478667 | 0.230822 | Biso | 1.000000 H |
| H24 | 1.0 | 0.496182 | 0.478667 | 0.730822 | Biso | 1.000000 H |
| H25 | 1.0 | 0.252509 | 0.250000 | 0.507519 | Biso | 1.000000 H |
| H26 | 1.0 | 0.000519 | 0.250000 | 0.753488 | Biso | 1.000000 H |
| H27 | 1.0 | 0.499481 | 0.750000 | 0.253488 | Biso | 1.000000 H |
| H28 | 1.0 | 0.247491 | 0.750000 | 0.007519 | Biso | 1.000000 H |
| H29 | 1.0 | 0.999481 | 0.250000 | 0.253488 | Biso | 1.000000 H |
| H30 | 1.0 | 0.747491 | 0.250000 | 0.007519 | Biso | 1.000000 H |
| H31 | 1.0 | 0.752509 | 0.750000 | 0.507519 | Biso | 1.000000 H |
| H32 | 1.0 | 0.500519 | 0.750000 | 0.753488 | Biso | 1.000000 H |

(d) Ima2 structure

#======================================================================

# CRYSTAL DATA

#----------------------------------------------------------------------

data_VESTA_phase_1

_pd_phase_name                    'f————————— S   H              '
_cell_length_a                    6.17800
_cell_length_b                    6.17800
_cell_length_c                    6.17800
_cell_angle_alpha                   90
_cell_angle_beta                    90
_cell_angle_gamma                   90
_symmetry_space_group_name_H-M    'P 1'
_symmetry_Int_Tables_number         1

loop_
_symmetry_equiv_pos_as_xyz
 'x, y, z'

loop_
  _atom_site_label
  _atom_site_occupancy
  _atom_site_fract_x
  _atom_site_fract_y
  _atom_site_fract_z
  _atom_site_adp_type
  _atom_site_B_iso_or_equiv
  _atom_site_type_symbol
   S1    1.0   0.000000   0.000000   0.000000   Biso  1.000000 S
   S2    1.0   0.000000   0.000000   0.500000   Biso  1.000000 S
   S3    1.0   0.000000   0.500000   0.000000   Biso  1.000000 S
   S4    1.0   0.000000   0.500000   0.500000   Biso  1.000000 S
   S5    1.0   0.500000   0.000000   0.000000   Biso  1.000000 S
   S6    1.0   0.500000   0.000000   0.500000   Biso  1.000000 S
   S7    1.0   0.500000   0.500000   0.000000   Biso  1.000000 S
   S8    1.0   0.500000   0.500000   0.500000   Biso  1.000000 S
   S9    1.0   0.250000   0.250000   0.250000   Biso  1.000000 S
   S10   1.0   0.250000   0.250000   0.750000   Biso  1.000000 S
   S11   1.0   0.250000   0.750000   0.250000   Biso  1.000000 S
   S12   1.0   0.250000   0.750000   0.750000   Biso  1.000000 S

| | | | | | | |
|---|---|---|---|---|---|---|
| S13 | 1.0 | 0.750000 | 0.250000 | 0.250000 | Biso | 1.000000 S |
| S14 | 1.0 | 0.750000 | 0.250000 | 0.750000 | Biso | 1.000000 S |
| S15 | 1.0 | 0.750000 | 0.750000 | 0.250000 | Biso | 1.000000 S |
| S16 | 1.0 | 0.750000 | 0.750000 | 0.750000 | Biso | 1.000000 S |
| H1 | 1.0 | 0.267044 | 0.021028 | 0.007515 | Biso | 1.000000 H |
| H2 | 1.0 | 0.267044 | 0.978972 | 0.507515 | Biso | 1.000000 H |
| H3 | 1.0 | 0.267044 | 0.478972 | 0.007515 | Biso | 1.000000 H |
| H4 | 1.0 | 0.267044 | 0.521028 | 0.507515 | Biso | 1.000000 H |
| H5 | 1.0 | 0.767044 | 0.978972 | 0.007515 | Biso | 1.000000 H |
| H6 | 1.0 | 0.767044 | 0.021028 | 0.507515 | Biso | 1.000000 H |
| H7 | 1.0 | 0.767044 | 0.521028 | 0.007515 | Biso | 1.000000 H |
| H8 | 1.0 | 0.767044 | 0.478972 | 0.507515 | Biso | 1.000000 H |
| H9 | 1.0 | 0.019347 | 0.250000 | 0.016936 | Biso | 1.000000 H |
| H10 | 1.0 | 0.983064 | 0.250000 | 0.480653 | Biso | 1.000000 H |
| H11 | 1.0 | 0.983064 | 0.750000 | 0.980653 | Biso | 1.000000 H |
| H12 | 1.0 | 0.019347 | 0.750000 | 0.516936 | Biso | 1.000000 H |
| H13 | 1.0 | 0.483064 | 0.250000 | 0.980653 | Biso | 1.000000 H |
| H14 | 1.0 | 0.519347 | 0.250000 | 0.516936 | Biso | 1.000000 H |
| H15 | 1.0 | 0.519347 | 0.750000 | 0.016936 | Biso | 1.000000 H |
| H16 | 1.0 | 0.483064 | 0.750000 | 0.480653 | Biso | 1.000000 H |
| H17 | 1.0 | 0.992485 | 0.021028 | 0.232956 | Biso | 1.000000 H |
| H18 | 1.0 | 0.992485 | 0.978972 | 0.732956 | Biso | 1.000000 H |
| H19 | 1.0 | 0.992485 | 0.478972 | 0.232956 | Biso | 1.000000 H |
| H20 | 1.0 | 0.992485 | 0.521028 | 0.732956 | Biso | 1.000000 H |
| H21 | 1.0 | 0.492485 | 0.978972 | 0.232956 | Biso | 1.000000 H |
| H22 | 1.0 | 0.492485 | 0.021028 | 0.732956 | Biso | 1.000000 H |
| H23 | 1.0 | 0.492485 | 0.521028 | 0.232956 | Biso | 1.000000 H |
| H24 | 1.0 | 0.492485 | 0.478972 | 0.732956 | Biso | 1.000000 H |
| H25 | 1.0 | 0.247908 | 0.250000 | 0.508488 | Biso | 1.000000 H |
| H26 | 1.0 | 0.991512 | 0.250000 | 0.752092 | Biso | 1.000000 H |
| H27 | 1.0 | 0.991512 | 0.750000 | 0.252092 | Biso | 1.000000 H |
| H28 | 1.0 | 0.247908 | 0.750000 | 0.008488 | Biso | 1.000000 H |
| H29 | 1.0 | 0.491512 | 0.250000 | 0.252092 | Biso | 1.000000 H |
| H30 | 1.0 | 0.747908 | 0.250000 | 0.008488 | Biso | 1.000000 H |
| H31 | 1.0 | 0.747908 | 0.750000 | 0.508488 | Biso | 1.000000 H |
| H32 | 1.0 | 0.491512 | 0.750000 | 0.752092 | Biso | 1.000000 H |

[5] Fermi surface of the Ima2 structure of $(SH^-)(H_3S^+)$

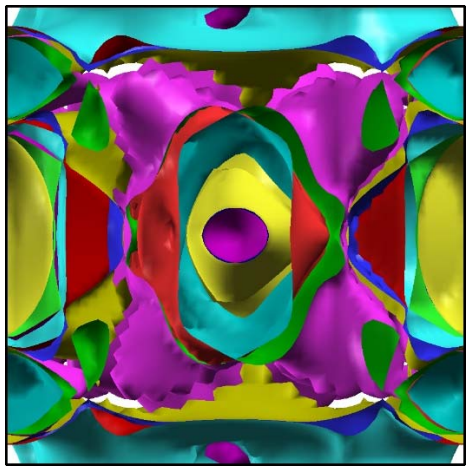 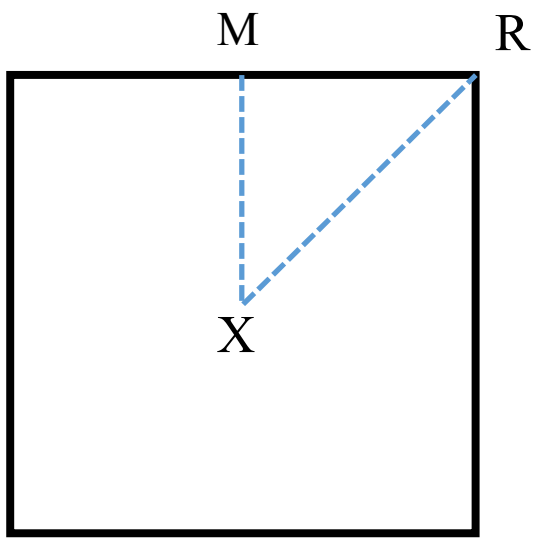

**Corrigendum for** E. E. Gordon et al, *Angew. Chem. Int. Ed*. **2016**, *55*, 3682.

In our estimation of $\Delta H_1$ for the reaction,

$$2H_3S \rightarrow (SH^-)(H_3S^+) + H_2, \qquad (1)$$

we made an inadvertent error of using half the energy of $H_2$ instead of the full energy. For the *P4mm*, *Cmmm*, *Cmc*21 and *Ima*2 structures of $(SH^-)(H_3S^+)$, the correct $\Delta H_1$ values are -3.200, -3.511, -3.559 and -3.582 eV, respectively. Thus, at 200 K, the correct $\Delta G_1$ values are -3.219, -3.529, -3.578 and -3.601 eV, respectively. These estimates are valid at ambient pressure because we employed the energy of $H_2$ calculated for its structure at ambient pressure.

To estimate the $\Delta H_1$ values at ultrahigh pressure P, it is necessary to consider the VP term, where V is the cell volume. In our calculations, the $Im\bar{3}m$ structure of $H_3S$ and the structures of $(SH^-)(H_3S^+)$ have the same cell volume, so their relative energies are not affected by the VP term. However, one needs to consider the VP term for $H_2$. Under ultrahigh pressure, it is most likely that $H_2$ dissociates into H atoms, and the latter are present on the surface or at the interstitial sites of the Re metal (i.e., the gasket material for the high-pressure cell) forming Re-H bonds. Then, the VP term for "$H_2$" vanishes. It is reasonable to assume that the loss of H-H bonding is compensated by the Re-H bonding. Then, to a first approximation, our estimates of $\Delta H_1$ at ambient pressure are also reasonable at ultrahigh pressure.

For the following reaction,

$$1.5(SH^-)(H_3S^+) \rightarrow 2H_3S + S, \qquad (2)$$

where S refers to the β-Po structure of elemental sulfur, our calculations show that the enthalpy changes $\Delta H_2$, under the pressure of 160 GPa, for the *P4mm*, *Cmmm*, *Cmc*21 and *Ima*2 structures of $(SH^-)(H_3S^+)$ are -3.630, -3.163, -3.091 and -3.056 eV, respectively. The activation energies of both reactions, Eq. 1 and Eq. 2, cannot be both small. Otherwise, all $H_2S$ under ultrahigh pressure would be converted to elemental S in disagreement with experiment. Due to the possible catalytic effect of the Re metal surface, Eq. 1 is more likely to have a low activation than does Eq. 2.

We thank Dr. Igor Mazin and Dr. Noam Bernstein for alerting us to recheck our calculations.